\documentclass[aip,apl,amsmath,amssymb,reprint,showpacs,showkeys,groupedaddress]{revtex4-1}

\usepackage{graphicx}
\usepackage{dcolumn}

\begin{document}

\title{Electronic structure of InAs quantum dots with GaAsSb strain reducing layer: localization of holes and its effect on the optical properties}

\author{P. Klenovsk\'y}
\email{klenovsky@physics.muni.cz}
\author{V. K\v{r}\'apek}
\author{D. Munzar}
\author{J. Huml\'i\v{c}ek}
\affiliation{
Department of Condensed Matter Physics, Faculty of Science, Masaryk University,
Kotl\'a\v{r}sk\'a~2, 61137~Brno, Czech~Republic}

\date{\today}

\begin{abstract}
The electronic structure of InAs quantum dots covered with the $\mathrm{GaAs_{1-y}Sb_y}$ strain reducing layer has been studied using the $\vec{k}\cdot\vec{p}$ theory. We explain previous experimental observations of the red shift of the photoluminescence emission with increasing $y$ and its blue shift with increasing excitation power. For $y>0.19$ type-II dots are formed with holes localized in the GaAsSb close to the dot base; two segments at opposite sides of the dot, forming molecular-like states, result from the piezoelectric field. We also propose an experiment that could be used to identify the hole localization using a vertical electric field.

\end{abstract}



\maketitle

InAs quantum dots (QDs) covered with the $\mathrm{GaAs_{1-y}Sb_y}$
strain reducing layer (SRL) represent an efficient light source
at the communication wavelengths of
$1.3/1.55\,\mathrm{\mu m}$.~\cite{Liu,LiuSteer}
The role of the SRL is to reduce the
strain inside the QDs and to modify the confinement potential, both effects increasing the otherwise small emission wavelengths of the InAs
QDs. Note that the $\mathrm{GaAs_{1-y}Sb_y}$ capped QDs differ from the more conventional $\mathrm{In_xGa_{1-x}As}$ capped ones~\cite{Mlinar2,Chen} in several important aspects. 
The antimony atoms act as surfactant,
helping to preserve the shape and the height of the QDs during the overgrowth. 
The resulting QDs are considerably higher than the $\mathrm{In_xGa_{1-x}As}$-capped ones.~\cite{Ulloa}
Moreover, for $y>0.14$, the formation of type-II dots with holes localized
in the SRL was reported.~\cite{Jin, Liu, LiuSteer} 
The red shift of the emission energy with $y$ increasing from $0.10$ to $0.22$
has been found,~\cite{LiuSteer} more pronounced in the type-II confinement
regime. A peculiarity of the type-II regime is a large blue shift (up to tens of meV)
of the emission energy with the excitation power.~\cite{Jin, LiuSteer}

We have calculated the electronic structure of QDs capped by the $\mathrm{GaAs_{1-y}Sb_y}$ SRL using a two-step approach.
First, single particle states were obtained using the Nextnano++ solver~\cite{next} 
based on the eight-band
$\vec{k}\cdot\vec{p}$ theory.
Second, we calculated the energies of the excitons using the configuration
interaction method with the basis set formed from the single particle
states.~\cite{Rodt} The values of the material
constants are given in Ref.~\onlinecite{sup1}. 
Based on published cross-section scanning tunneling micrographs (XSTM),~\cite{Ulloa,Offermans}
we assume a pyramidal shape of the model QD with the height of
8~nm and the base length of 22~nm. Both the shape and the dimensions were taken from Ref.~\onlinecite{Ulloa} without any changes.
The results of Refs.~\onlinecite{Ulloa} and~\onlinecite{Offermans} also
suggest that the QDs are composed of $\mathrm{In_{x}Ga_{1-x}As}$
with a trumpet-like composition profile, 
with $x=0.4$ at the base of the pyramid (low In content) and
$x=1.0$ at its top (maximum In content). Here we adopt this profile. The XSTM cannot determine the composition profile very reliably,~\cite{Mlinar1} thus we studied several other profiles and found only minor differences.~\cite{sup2} 
The thickness of the SRL
has been fixed at 5~nm and $y$ varied between 0.10 and 0.22. 

\begin{figure}[ht]
\includegraphics{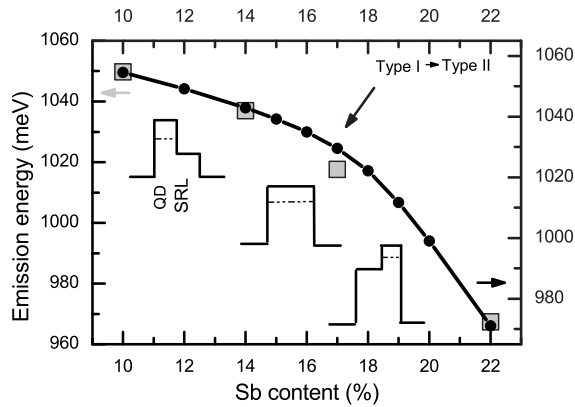}
\caption{
Calculated values of the emission energy of the model pyramidal QD 
(circles)
compared to the corresponding values obtained from the optical measurements, taken from Ref.~\onlinecite{LiuSteer}
(squares). The vertical axis of the experimental values is shifted by 5~meV with respect to that of the model results.
The arrow points to the onset of the type-II confinement. 
A schematic representation of the hole confinement is also shown.
\label{fig1}}
\end{figure}

Figure~\ref{fig1} shows the lowest transition energy as a function of $y$. With $y$ increasing
from 0.10 to 0.22, the energy red-shifts by 84~meV from 1055~meV to 971~meV, the origin being illustrated in the inset. 
This is in very good agreement with the optical data,~\cite{LiuSteer}
where the energy shifts by 83~meV from 1050~meV to 967~meV. 
We emphasize that the red shift was reproduced without any modification
of the shape and dimensions of the model QD.
This is consistent with the observations of Ref.~\onlinecite{Ulloa} that the shape and size of the QDs are independent of $y$. 
Recall that in the $\mathrm{In_{x}Ga_{1-x}As}$ capped QDs, a large part of the red shift is caused by the increase of volume.~\cite{Kuldova2006} 
For $y<0.15$ ($y>0.19$), the probability of finding the hole inside the QD is larger than $80\,\%$ (smaller than $20\,\%$), which we call type-I (type-II) confinement in the following.
We have found the type I to type II transition not to be abrupt, but rather to be a continuous process starting at low $y$, accelerated in a certain range of $y$ ($0.15<y<0.19$ for our dot), 
and finally saturating at high values of $y$.~\cite{sup3} Note that we found the type-II QDs for larger $y$ than usually reported. 
The redshift of the type-II confinement is faster because the bottom of the potential well shifts and not the barrier as in type I.~\cite{Liu,t_mlinar,Klenovsky2010}

The type I to type II transition is associated with a reduction of the overlap of the electron and hole wave functions and the corresponding decrease of the Coulomb energy of the electron-hole pair from 26~meV at $y=0.10$ to 12~meV at $y=0.22$. 
The biexciton emission energy calculated using the configuration interaction method~\cite{Rodt} with
6 electron and 8 hole basis states is blue-shifted
by 4~meV with respect to the exciton emission energy at $y=0.10$,
and blue-shifted by 18~meV at $y=0.22$. 
While the multiexcitonic ordering in type-I QDs depends on structural properties of the QD~\cite{Mlinar1}, the blue shift of the biexciton with respect to the exciton is a characteristic property of the type-II QDs.~\cite{sup2} We suggest
that the blue shift of the emission energy with the
pumping intensity reported for type-II QDs~\cite{Jin, LiuSteer} is caused
by the emission from the blue-shifted biexcitons emerging at high pumping powers.
This could be verified by intensity-dependent
micro-photoluminescence measurements. 
The biexciton energy could be used as the indicator of the type of the confinement.

\begin{figure}[ht]
\includegraphics{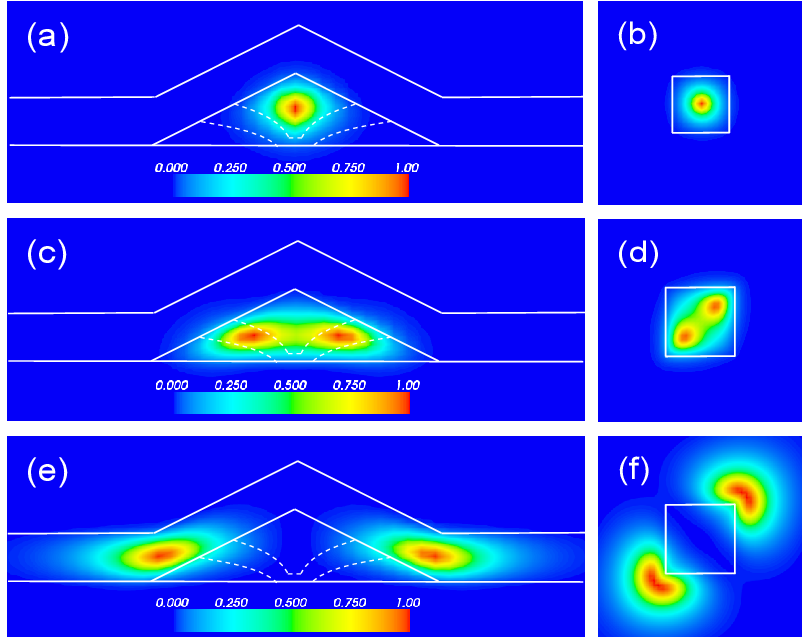}
\caption{
Cross sections of the probability density (arbitrary units) of the lowest
electron state for $y=0.10$: (a) ($1\bar{1}0$) plane,
(b) ($001$) plane; (c) and (d): the same for the lowest hole state and $y=0.10$; (e) and (f): the same for the lowest hole state and $y=0.22$ (see also Ref.~\onlinecite{sup4}). 
The ($1\bar{1}0$) planes contain the vertical symmetry axis of the QD and the ($001$) cross sections are located 3.5~nm (2.5~nm) above the base of the QD for electrons (for holes).
The borders of the QD and the SRL are denoted by the solid lines.
The dashed lines are the isolines of the In content inside the QD
(100 and 60~\%).
\label{fig2}}
\end{figure}

\begin{figure}[ht]
\includegraphics{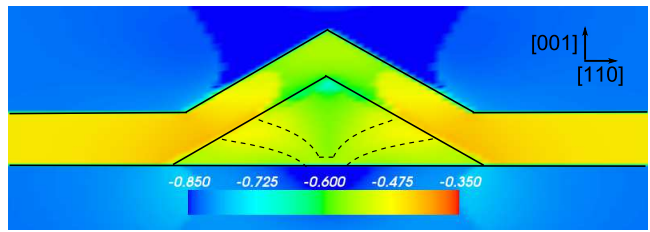}
\caption{
The ($1\bar{1}0$) cross section of the confinement potential for holes in the type-II QD with $y=0.22$.
The scale is in eV. The holes are confined in the two potential minima
in the SRL close to the base of the QD.
\label{fig3}}
\end{figure}

Next we address the wave functions.
The probability densities of the lowest electron and hole states in the model QD are shown
in Fig.~\ref{fig2} ($y=0.10$ for electrons, $y=0.10$ and $y=0.22$ for holes).
The electrons are localized in the In-rich region of the QD, the position of the maximum of their probability density being only weakly dependent on $y$. 
For $y=0.10$, the holes are localized inside the QD close to the
base of the pyramid, in agreement with previously reported results for the type-I InAs QDs.~\cite{Stier1999}
For $y>0.19$, the holes are
located in the SRL. It has been assumed that they reside above the QD.~\cite{Jin, Alice} We found them, however, to be located close to the base of the dot (see Figs.~\ref{fig2}e,\,f). This is a consequence
of the strain distribution. The shift of the heavy-hole band edge due to the biaxial strain $B=\varepsilon_{zz}-(\varepsilon_{xx}+\varepsilon_{yy})/2$ is given by $\Delta E=-b \cdot B$, where $\varepsilon_{ij}$ are the components of the strain tensor and $b$ is the biaxial deformation potential. Since $b$ is negative, the holes prefer the regions with large positive values of $B$ at the dot base as shown in~Fig.~\ref{fig3}.

If only the strain is included in the calculations, the maximum of the probability density of the hole ground state forms a ring surrounding the dot base.
The piezoelectric field splits it into two segments located along the [110] direction.~\cite{sup4}
The wave functions resemble those of quantum dot molecules. 
Quantum dot molecules (QDM) are important for the quantum information
processing as a potential tool for manipulating the qubits -- a quantum
gate.~\cite{Burkard}
Vertical QDMs can be easily manufactured and
offer a strong coupling, but they cannot be scaled
and suffer from the internal symmetry lowering induced by the
strain field.~\cite{Bester}
On the other hand, lateral QDMs (LQDM) reported so far exhibit a weak coupling~\cite{Beirne}
and their properties cannot be tuned easily. 
The properties of the ``molecules" under
consideration are in many aspects superior.
Most importantly, the distance between the segments and
the tunneling energy
can be easily tuned by the size of the QD and the Sb content in the SRL 
(the value of the tunneling energy corresponding to the QDs 
under consideration is 0.28 meV). 
The barrier between the segments is better defined. 
The clear disadvantage is the confinement of only one type
of charge carriers, i.e., holes, 
ruling out applications in devices involving both electrons and holes. 
However,
there exist promising proposals of devices which require only one type
of charge carriers.~\cite{Burkard} 
Our ``molecules" represent an interesting alternative to the LQDMs discussed in this context.

\begin{figure}[ht]
\includegraphics{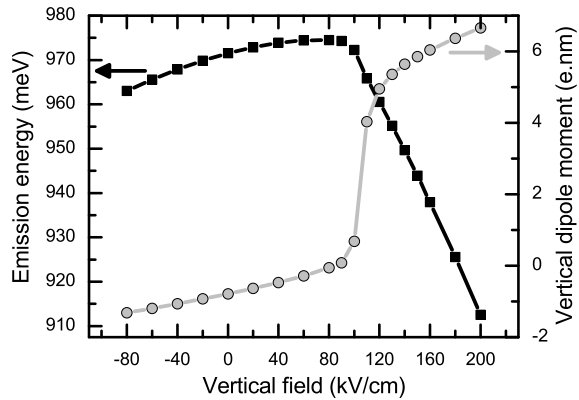}
\caption{
The emission energy (squares) and the vertical dipole moment
(circles) as functions of the vertical electric field in the type-II QD with $y=0.22$.
\label{fig4}}
\end{figure}

In the following we propose an experiment that could be used to identify the localization of the holes,
based on the Stark shift in the vertical electric field.
Figure~\ref{fig4} shows the calculated transition energy and the vertical dipole moment
as functions of the applied electric field. 
According to our calculations,
the dipole moment of the exciton in zero field is small and oriented downwards (negative direction).
The electron and the hole are both rather strongly confined and their
vertical polarizability is low. As a consequence, only a weak Stark shift
of the emission energy in a vertical electric field is predicted,
positive when the field is oriented upwards (positive direction).
For a sufficiently strong positive field, however, the hole state
above the tip of the QD turns to be the ground state,~\cite{sup5} 
the dipole moment of the exciton increases dramatically and
changes its sign, and so does the Stark shift.
The transition between the weak negative
and large positive dipole moment is fairly steep and should be identifiable. For the hole
ground state located above the tip of the QD, a similar transition would occur
for a {\it negative} field. 
We note that this possibility could occur in real QDs due to, e.g.,
a higher antimony content in the SRL above the QD.~\cite{Ulloa2}

In conclusion, we have shown that the red shift of the emission energy
with increasing Sb content $y$ in the $\mathrm{GaAs_{1-y}Sb_y}$ SRL is mainly due to the modified confinement potential in the QD;
there is no need to vary the structural parameters of the model dot with changing $y$ to reproduce the optical data. The blue shift of the emission with increasing pumping power is attributed to the increasing proportion of the biexcitons whose energy is considerably higher compared to the excitons.
We have found that for $y>0.19$ the holes are localized in the SRL close to the base of the dot, rather than above it.
An experiment involving external electric field has been proposed to verify this prediction.

The work was supported by the Institutional research program
MSM~0021622410 and the GACR grant GA202/09/0676.

%



\end{document}